\pdfoutput=1

\documentclass[
  aps,
  prd,
  10pt,
  superscriptaddress,
  amsmath,
  amssymb,
  twocolumn,
  nofootinbib,
  nobibnotes
]{revtex4-2}
\usepackage{dsfont}

\usepackage{graphicx}
\usepackage{dcolumn}
\usepackage{bm}
\usepackage{xcolor}
\usepackage{xfrac}
\usepackage{physics}
\usepackage{hyperref}
\usepackage{bbm}
\usepackage{mathrsfs}
\usepackage[utf8]{inputenc}
\usepackage{cancel}
\usepackage{amsmath}
\usepackage[normalem]{ulem}

\usepackage{mathtools}

\usepackage{orcidlink}

\renewcommand{\(}{\left(}
\renewcommand{\)}{\right)}
\renewcommand{\[}{\left[}
\renewcommand{\]}{\right]}

\usepackage{tcolorbox}

\let\originalleft\left
\let\originalright\right
\renewcommand{\left}{\mathopen{}\mathclose\bgroup\originalleft}
\renewcommand{\right}{\aftergroup\egroup\originalright}

\newcommand{\vect}[1]{\boldsymbol{#1}}
\renewcommand{\vec}[1]{\vect{#1}}

\makeatletter
\newcommand*\bigcdot{{\color{gray}\mathpalette\bigcdot@{1.}}}
\newcommand*\bigcdot@[2]{\mathbin{\vcenter{\hbox{\scalebox{#2}{$\m@th#1\bullet$}}}}}
\makeatother

\def\rme{{\mathrm {e}}}
\def\rmi{{\mathrm {i}}}
\newcommand{\rmd}{\mathrm{d}}

\newcommand{\idhat}{\hat{\mathds{1}}}

\newcommand{\delhat}{\hat{\delta}}

\newcommand{\rhohat}{\hat{\rho}}
\newcommand{\rhotil}{\tilde{\rho}}

\newcommand{\sigmam}{\hat{\sigma}^{-}}

\newcommand{\sigmax}{\hat{\sigma}^{x}}

\newcommand{\sigmahat}{\hat{\sigma}}

\newcommand{\chivec}{{\vec{\chi}}}

\newcommand{\Ahat}{\hat{A}}

\newcommand{\Bhat}{\hat{B}}

\newcommand{\cscr}{\mathscr{C}}

\newcommand{\dcal}{\mathcal{D}}

\newcommand{\Lhat}{\hat{L}}
\newcommand{\dLhat}{\hat{L}^\dagger}

\newcommand{\Lcal}{\mathcal{L}}

\newcommand{\nvec}{{\vec{n}}}

\newcommand{\Ohat}{\hat{O}}

\newcommand{\bea}{\begin{equation}\begin{aligned}}
		\newcommand{\eea}{\end{aligned}\end{equation}}
\newcommand{\be}{\begin{equation}}
	\newcommand{\ee}{\end{equation}}

\def\bra#1{\langle#1 |}
\def\ket#1{| #1\rangle}

\def\ee{\mathrm{e}}

\renewcommand{\Tr}{\mathrm{Tr}}

\newcommand{\PauliSigma}{\hat{\sigma}}

\newcommand{\calN}{\mathcal{N}}

\newcommand{\Ham}{\widehat{H}}

\usepackage{cleveref}

\newcommand{\oprho}{\hat{\rho}}

\begin{document}

\preprint{APS/123-QED}

\title{Quantum jump correlations in long-range dissipative spin systems via cluster and cumulant expansions}

\author{Giulia Salatino~\orcidlink{0009-0005-6913-6920}}
\thanks{These authors contributed equally to this work.}
\address{Scuola Superiore Meridionale, Via Mezzocannone 4, 80138 Napoli, Italy}
\author{Anna Delmonte~\orcidlink{0009-0008-9371-6855}}
\thanks{These authors contributed equally to this work.}
\address{CNRS, Collège de France, 11 place Marcelin-Berthelot, 75321 Paris Cedex 05, France}
\affiliation{SISSA, Via Bonomea 265, I-34136 Trieste, Italy}
\author{Zejian Li~\orcidlink{0000-0002-5652-7034}}
\thanks{These authors contributed equally to this work.}
\affiliation{The Abdus Salam International Centre for Theoretical Physics (ICTP), Strada Costiera 11, 34151 Trieste, Italy}
\address{Université Paris Cité, CNRS, Matériaux et Phénomènes Quantiques, 75013 Paris, France}
\author{Rosario Fazio~\orcidlink{0000-0002-7793-179X}}
\affiliation{The Abdus Salam International Centre for Theoretical Physics (ICTP), Strada Costiera 11, 34151 Trieste, Italy}
\affiliation{Dipartimento di Fisica “E. Pancini”, Università di Napoli “Federico II”, Monte S. Angelo, I-80126 Napoli, Italy}
\author{Alberto Biella~\orcidlink{0000-0001-9301-1638}}
\email{alberto.biella@cnr.it}
\affiliation{Pitaevskii BEC Center, CNR-INO and Dipartimento di Fisica, Università di Trento, I-38123 Trento, Italy}
\affiliation{INFN-TIFPA, Trento Institute for Fundamental Physics and Applications, I-38123 Trento, Italy}

\date{\today}

\begin{abstract}
We characterize nonequilibrium phases in long-range dissipative spin systems through the statistical properties of quantum jump trajectories. 
While the average dynamics governed by the Lindblad master equation provides access to steady-state expectation values of order parameters, the quantum trajectory framework reveals features encoded in the spatial and temporal correlations of detection events.
Focusing on a model exhibiting a paramagnetic-to-ferromagnetic phase transition, we investigate the full counting statistics of quantum jumps using a tilted Lindbladian approach. 
We combine this with cluster mean-field and cumulant expansion techniques, which allow us to capture, respectively, the short- and long-range structure of jump correlations. 
In addition, we study the waiting-time distributions of detection events. 
We show that quantum jump correlations display clear signatures of the underlying phases and reveal distinct dynamical features across the transition. 
Our results highlight the potential of trajectory-resolved observables as probes of collective behavior in open quantum many-body systems and provide new insights into the role of long-range interactions in shaping nonequilibrium dynamics.
\end{abstract}

\maketitle

\section{Introduction}

Understanding the emergence of collective behavior in open quantum systems is a central challenge in many-body physics~\cite{fazio2025manybodyopenquantumsystems,MingantiBiella2026}. 
In particular, dissipative spin systems are paradigmatic platforms where the interplay between coherent interactions and environmental coupling gives rise to rich nonequilibrium phenomena. 
These systems can exhibit collective phases and phase transitions characterized by spontaneous symmetry breaking, even in the absence of thermal equilibrium (see, e.g., Refs.~\cite{Lee2011,Chan2015,Landa2020,Biella2018,Jin2018_02}). 
Long-range interactions further enrich this landscape, enabling the stabilization of ordered phases also in one spatial dimension~\cite{paz2024entanglement,defenuOutofequilibriumDynamicsQuantum2024}.

The characterization of such nonequilibrium phases usually relies on the analysis of steady-state properties. 
In this framework, one typically identifies suitable order parameters (such as magnetization) and studies their expectation values in the long-time limit of the dynamics, described by a Lindblad master equation. 
By tracking how these observables behave as a function of system parameters, one can infer the presence of phase transitions and classify the corresponding phases~\cite{Minganti2018,Kessler2012}. 

Another perspective emerges when the dissipative evolution is interpreted as the ensemble average over stochastic quantum trajectories associated with continuous monitoring of the environment degrees of freedom~\cite{molmer1993monte,Daley04032014}. 
In this picture, individual realizations of the dynamics are punctuated by quantum jumps, corresponding to detection events (or {\it clicks}) in the measurement record~\cite{landi2024current}: the position in space and time of the detection events depends from the many-body wavefunction and can carry detailed information about the system’s collective state.

This idea led to the so-called thermodynamics of trajectories~\cite{garrahan2010thermodynamics, carollo2023thermoquantcomp}, where dynamical fluctuations are probed by biasing the probability of trajectories according to time-integrated observables, such as the dynamical activity.
In the quantum-jump setting considered below, this biased evolution takes the form of a {\it tilted} Lindblad master equation, where the counting field weights the jump term and thereby provides access to the full counting statistics of detection events. 
This approach has been extensively used to identify dynamical phase transitions in a variety of settings, including kinetically constrained models~\cite{vanwijland2007constrained, garrahan2012facilitated, carollo2025constraineddiscretetime, cea2026largedeviationsconditionedmonitored, cech2026classicalstochasticmonitoredquantum}, superconducting circuits~\cite{garrahan2012super}, Rydberg gases~\cite{garrahan2018rydberg}, and interacting spin models~\cite{garrahan2012ising,garrahan2013manymodels,fitzner2026}. 
More recently, correlations between quantum jump events have been shown to be linked to changes in the entanglement structure of monitored quantum systems~\cite{yamamoto2026_01,yamamoto2026_02}.

In open systems with long-range interactions~\cite{liMonitoredLongrangeInteracting2025a,delmonteMeasurementinducedPhaseTransitions2025,liEmergentDeterministicEntanglement2025, lesanovsky2024timecrystals}, the nature of these correlations remains largely unexplored. 
In particular, due to the non-ballistic spread of excitations and entanglement~\cite{Eisert2013,Hauke2013,Schachenmayer,Santos2016,Cevolani2018,Schneider2021}, it is not yet clear how spatial and temporal correlations between detection events encode the presence of different phases, nor how they are influenced by the collective nature of the dynamics.

Our work builds on this trajectory-based viewpoint, addressing these questions focusing on a long-range interacting monitored Ising chain and on quantities directly tied to the spatial organization of detection events. 
The considered model exhibits a transition between paramagnetic and ferromagnetic phases as a function of interaction and dissipation strength~\cite{sierant2022dissipative}. 
Rather than analyzing the large-deviation properties of the total dynamical activity, we use the tilted-Lindbladian formalism to study the full counting statistics of site-resolved quantum jumps. In particular, we focus on correlations between the numbers of jumps occurring on different sites, combining cluster mean-field theory~\cite{russo2025quantum,JinPRX2016,Kikuchi1951,Bethe1935} and cumulant expansions~\cite{Verstraelen2023,FRICKE1996479,Kubo1962} to resolve short-range and long-range structures in the measurement record. 
This framework enables us not only to characterize the emergence of nonlocal jump correlations, but also to analyze waiting-time distributions associated with the detection record.

Within the cluster mean-field approach, we find that the ferromagnetic phase is marked by a joint probability distribution of jumps on neighboring sites that broadens in time, while its connected part develops a characteristic four-quadrant structure, signaling their anti-correlation. By contrast, in the paramagnetic phase the joint probability distribution becomes much weaker and approximately factorized. The same trend is reflected in the growth rate of the covariance of the number of jumps, which is finite and negative in the ordered phase and tends to vanish in the disordered regime.

Within the same framework, we also analyze the waiting-time distribution between consecutive jumps. At the single-site mean-field level, we derive an analytical expression for this quantity and show that its mean and variance remain finite in the ferromagnetic phase, whereas they diverge in the paramagnetic one because of the approach to a dark state. Incorporating short-range correlations through larger clusters preserves this qualitative distinction, while shifting the transition point and progressively reducing the region with finite waiting-time moments as the interaction range decreases.

To complement the cluster analysis, we develop a second-order cumulant expansion for the tilted Lindbladian, which captures long-range correlations and finite-size effects. In the infinite-range limit, jump correlations are independent of distance, are anti-correlated in the ferromagnetic phase, and change sign close to the transition before becoming weakly positive in the paramagnetic regime. For power-law interactions, nearest-neighbor correlations are strongly enhanced near criticality and display a pronounced system-size dependence, whereas at larger distances they recover a more mean-field-like behavior.

Altogether, these results show that full counting statistics and waiting-time distributions provide a trajectory-based characterization of dissipative phases in long-range spin systems. In this picture, cluster mean-field theory resolves the short-range structure of jump correlations, while the cumulant expansion captures their long-range and critical features.

The remainder of the paper is structured as follows. In Sec.~\ref{sec_model}, we introduce the model and the methods, including the long-range dissipative spin system, the quantum-jump trajectory picture, the tilted Lindbladian formalism, and the waiting-time distribution. In Sec.~\ref{sec_fcs}, we investigate the full counting statistics of quantum jumps using cluster mean-field theory and a cumulant expansion, emphasizing the complementary information provided by the two approaches. In Sec.~\ref{sec_wtd}, we turn to the waiting-time distribution and show how its behavior provides an additional characterization of the dissipative phases. Finally, in Sec.~\ref{sec_conclusions}, we present our conclusions.

\section{Model and methods}
\label{sec_model}

We focus on a long-range interacting spin model in one dimension, $D=1$, exhibiting a rich dissipative phase diagram due to 
the interplay of interactions with an external transverse field, and losses.
In particular, the Hamiltonian governing the unitary dynamics of this system is
\begin{equation}\label{eq_Ham}
    \hat H = \hat H_{\scriptscriptstyle\mathrm{int}}+\hat H_{\scriptscriptstyle\mathrm{ext}}= -\frac{J}{2\mathcal N_\alpha}\sum_{\substack{i < j=1}}^N \, \frac{1}{r_{ij}^\alpha}\,\hat\sigma_i^x\,\hat\sigma_j^x + \, h\sum_{i=1}^N\,\hat\sigma_i^z,
\end{equation}
with $\hat \sigma_i^{\mu}, \mu=x,y,z$ the Pauli matrices on site $i$, and $\hbar=1$.
The parameter $J$ governs the strength of the ferromagnetic interaction, $h$ is the external aligning field, and $\alpha$ represents the range of the interaction. The distance $r_{ij}$ between sites $i,j$, considering periodic boundary conditions (PBC), is defined as $r_{ij}=\text{min}\left\{|i-j|,N-|i-j|\right\}$. The extensivity of the Hamiltonian is guaranteed by the presence of $\mathcal N_\alpha=\frac{1}{N}\sum_{ij}r_{ij}^{-\alpha}$, the Kac normalization. 

Note that this kind of Hamiltonian is not only interesting from the theoretical statistical mechanics point of view, but is also closely connected to the physics realized by current experimental platforms, which include long-range settings with interactions decaying as a power law of the distance~\cite{monroe2021programmable,ritsch2013cold,lahaye2009physics,bohn2017cold,weimer2010rydberg,ferioli2023non}.

The dynamics generated by the Hamiltonian in Eq.~\eqref{eq_Ham} depends sensibly on the interaction range $\alpha$. In the case $\alpha=0$, it represents an infinite-range model with all-to-all interactions between spins and is exactly solvable through mean-field. The regime $0<\alpha<1$ represents strong-long-range interactions and $\alpha\gtrsim2$ describes short-range interactions~\cite{sierant2022dissipative,vzunkovivc2018dynamical}. 
Here we will mainly focus on the long-range regime.

Spins on every site are subject to spontaneous emission given by the jump operator
\begin{equation}\label{eq_jumps}
    \hat L_i = \sqrt{\gamma} \hat\sigma_i^-\text{, }\,i=1,...,N,
\end{equation}
with $\hat\sigma_j^\pm=\hat\sigma_j^x\pm i\hat\sigma_j^y$, and $\gamma$ being the rate of the emission.

\begin{figure}[h]
    \centering
    \includegraphics[width=.9\columnwidth]{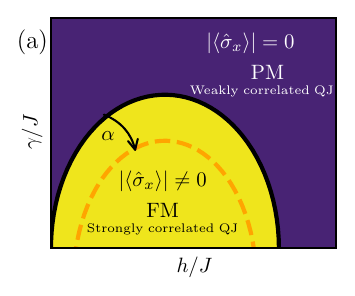}
    \includegraphics[width=.9\columnwidth]{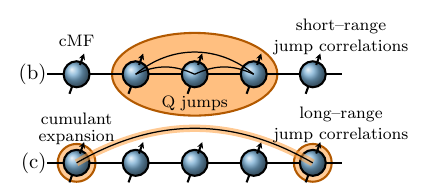}
    \caption{(a) Sketch of the dissipative phase diagram in the long-range limit: the yellow region corresponds to the mean-field ferromagnetic phase (FM), while the violet region corresponds to the mean-field paramagnetic phase (PM), featuring strongly and weakly correlated quantum jumps, respectively. The dashed line represents the shrinking of the FM phase due to an increased $\alpha$, reflecting the buildup of correlations between spins. Correlations between quantum jumps are captured through (b) a cluster mean-field (cMF) approximation, which fully accounts for short-range correlations, or (c) a cumulant expansion approximation, which captures lower-order long-range correlations.}
    \label{fig:single-mf}
\end{figure}

The dynamics of such a system is described by the action of the Lindbladian $\mathcal L$ on the quantum state of the system $\hat\rho$, namely
\begin{align}\label{eq_lindblad}
    \dot{\hat\rho} &=\mathcal L\hat\rho\\& =-i[\hat H,\hat\rho]+\gamma\sum_{j=1}^N\left(\hat\sigma_j^-\,\hat\rho\,\hat\sigma_j^+-\frac{1}{2}\left\{\hat\sigma_j^+\hat\sigma_j^-,\hat\rho\right\}\right)\nonumber.
\end{align}
The steady state of the above Lindblad equation reveals important properties of our long-range dissipative system. Depending on the range of the interaction, it is possible to witness a dissipative phase transition from a ferromagnetic phase with $|\expval{\hat \sigma_x}|>0$ to a paramagnetic phase with $|\expval{\hat \sigma_x}|=0$, as in Fig.~\ref{fig:single-mf} where the ferromagnetic (FM) region corresponds to the yellow phase, and the paramagnetic (PM) one to the violet region.

The ferromagnetic region is determined by the dominance of the $xx$-interaction term in the Hamiltonian, which tends to align spins within a range set by $\alpha$, while the paramagnetic region is related to the external field $h$, which tends to align spins along the $+z$ direction. The phase transition arises when $\alpha<\alpha_C$. Above the critical range $\alpha_C$, the ferromagnetic phase melts, yielding a steady state that features $|\expval{\hat \sigma_x}|=0$ across the whole parameter space. In equilibrium thermal phase transitions~\cite{paz2024entanglement,dysonExistencePhasetransitionOnedimensional1969}, the critical range is predicted to be $\alpha_C=1+D$, which is consistent with our numerical evidence in App.~\ref{app_dpt}, where we observe the melting of the phase transition for $\alpha>2$ through a cluster mean-field analysis, which serves as a baseline for comparing our new characterizations of the dissipative phase transitions via jump statistics.

\subsection{Jump statistics}
We now suppose to monitor the quantum jumps resulting from spontaneous emission of spins, and describe the dynamics through the quantum trajectories associated with the measurement outcome of such monitoring. The aim is then to use solely statistical information obtained from the monitoring of the system to describe features of the dissipative phase diagram, instead of standard order parameters as $|\expval{\sigma_x^i}|$ from the previous Section.

Every quantum trajectory is defined by two variables: the set of times at which spontaneous emission has been detected, and the sites of the spin from which the excitation has been emitted. We can thus represent a quantum trajectory of the evolution up to time $t$ as:
\begin{equation}\label{eq_traj}
    \Gamma(t): \left\{(t_1,i_1),(t_2,i_2),...,(t_n,i_n)\right\}\text{, }t_n<t.
\end{equation}
Knowing a trajectory allows to reconstruct the whole evolution of an initial pure state $\ket{\psi_0}$ conditioned on that measurement outcome as
\begin{align}\label{eq_traj}
    \ket{\tilde\psi_\Gamma(t)}\propto e^{-i\hat H_{\mathrm{nh}}(t-t_n)}\,\hat\sigma_{i_n}^-\,....&\,\hat \sigma_{i_2}^-\,e^{-i\hat H_{\mathrm{nh}}(t_2-t_1)}\nonumber\\&\,\hat\sigma_{i_1}^-\,e^{-i\hat H_{\mathrm{nh}}t_1}\ket{\psi_0},
\end{align}
where $\ket{\tilde\psi_\Gamma(t)}$ is non-normalized and 
\begin{equation}\label{eq_nhHam}
    \hat H_{\mathrm{nh}}=\hat H - i\frac{\gamma}{2}\sum_{j=1}^N\hat\sigma_j^+\hat\sigma_j^-
\end{equation}
is the non-Hermitian Hamiltonian corresponding to the smooth evolution without jumps~\cite{molmer1993monte,carmichael2008statistical,wiseman2009quantum,jacobs2014quantum}.

The dynamics of the average state, i.e. $\hat\rho=\frac{1}{N_{\rm{traj}}}\sum_\Gamma \ket{\psi_\Gamma}\bra{\psi_\Gamma}$ is simply given by the Lindblad equation in Eq.~\eqref{eq_lindblad}, where $\ket{\psi_\Gamma}$ corresponds to the normalized state of Eq.\eqref{eq_traj}. Averaging over trajectories discards information about the monitoring process, but preserves the dissipative effects generated by measuring the system and results in a Lindblad evolution.

Trajectories can be simulated in practice through a Monte Carlo wavefunction algorithm~\cite{molmer1993monte} and represent the building block of our work, enabling the calculation of several important quantities that describe the statistical properties of the jump dynamics. In the remaining part of the Section, we introduce these objects starting from the tilted Linbladian, used to describe the statistics of the number of jumps, and the waiting time distribution, which focuses on the statistics of time intervals between consecutive jumps.

\subsubsection{Tilted Linbladian}
The tilted Lindbladian $\mathcal L_{\chi}$ is connected to the study of the dynamics of the quantum state $\hat{\tilde\rho}_n(t)$ corresponding to evolution through $n$ jumps. 
In practice, the state $\hat{\tilde\rho}_n$ is obtained by averaging the quantum state over the ensemble of trajectories $\Gamma(t)$ which contain $n$ jumps up to time $t$:  $\hat{\tilde{\rho_n}}(t) = \frac{1}{N_{\mathrm{traj}|_n}}\sum_{\Gamma|_n}\ket{\tilde\psi_{\Gamma|_n}(t)}\bra{\tilde\psi_{\Gamma|_n}(t)}$, where $\bullet|_n$ indicates the restriction to trajectories with $n$ jumps and the trajectory states are not normalized.

The average state $\hat{\tilde\rho}_n$ follows the master equation~\cite{landi2024current}
\begin{align}
    &\hat{\tilde{\rho}}_n(t+dt) - \hat{\tilde{\rho}}_n(t) = -i \,dt\,\left[\hat H,\hat{\tilde\rho}_n(t)\right] \\&- \frac{\gamma}{2} \,dt\sum_{j=1}^N\left\{\hat \sigma_j^+\hat\sigma_j^-,\hat{\tilde\rho}_n(t)\right\} 
    +\gamma \, dt \, \sum_{j=1}^N\hat\sigma_j^-\,\hat{\tilde\rho}_{n-1}(t)\,\hat\sigma_j^+\nonumber.
\end{align}
Effectively, the evolution induced by the non-Hermitian Hamiltonian~\eqref{eq_nhHam}, corresponding to the commutator and anticommutator terms in the above equation, do not change the number of jumps of the state and contains only $\hat{\tilde\rho}_n$ itself. The last term, corresponding to quantum jumps, connects density matrices with a different number of jumps, in this case $\hat{\tilde\rho}_{n-1}$.
Defining the Fourier transform as
\begin{equation}
    \hat\rho_\chi = \sum_{n\in\mathbb N} e^{in\chi}\hat{\tilde\rho}_n\,,
\end{equation}
the non-trace-preserving master equation generated by the Tilted Lindbladian $\mathcal L_{\chi}$ reads for our system~\cite{landi2024current}:
\begin{align}\label{eq_tiltedL}
    \dot{\hat\rho}_\chi &=\mathcal L_\chi\hat\rho_\chi\\&= -i\left[\hat H,\hat\rho_\chi\right] +\gamma \sum_{j=1}^N\left(e^{i\chi}\hat \sigma_j^-\hat\rho_\chi\hat\sigma_j^+ -\frac{1}{2}\left\{\hat\sigma_j^+\hat\sigma_j^-,\hat\rho_\chi\right\}\right),\nonumber
\end{align}
where $\chi\in [0,2\pi)$ corresponds to the counting field and tilts the jump term only.
The probability of having $n$ jumps up to time $t$ is given by
\begin{equation}\label{eq:pn-fourier}
    P(n) = \Tr{\hat{\tilde\rho}_n(t)} = \int \frac{d\chi}{2\pi}\,e^{-in\chi}\, \Tr{\hat{\rho}_\chi(t)}.
\end{equation}
Thus, solving Eq.~\eqref{eq_tiltedL}, gives access to the full counting statistics of the number of jumps.

Note that the formalism can be generalized by counting the number of jumps on each site. In this case, we introduce the vector $\Vec{n}=(n_1,...,n_N)$ containing the number of jumps on each site, and the vector $\Vec{\chi}=(\chi_1,...,\chi_N)$ containing a counting field for every site. By appropriately shaping these vectors, all kinds of counting can be achieved, including cluster-counting  (for example, setting the same $\chi$ inside one cluster of spins).
We will be interested in all these objects in Sec.~\eqref{sec_fcs}.

\subsubsection{Waiting time distribution}
The last quantity we will consider is the \textit{waiting-time distribution} $\mathcal P(t)$. We focus our study of the waiting-time distribution on the case in which we monitor a single site $j$ of the spin chain, analyzing the statistics over trajectories of the time separating two consecutive jumps.

Formally, since we do not keep track of the quantum jumps on $N-1$ sites, and we let them evolve according to the Lindblad equation, we define a general no-click superoperator for the single monitored site as~\cite{landi2024current}
\begin{equation}
    \mathcal L_{{\rm{nh}}}^j = \mathcal L - \Lambda_j,
\end{equation}
where $\mathcal L$ is the many-body Lindbladian and $\Lambda_j\hat\rho=\gamma\,\hat \sigma_j^-\hat\rho\hat\sigma_j^+$.
The probability distribution of the waiting time between two consecutive jumps on site $j$ is then simply:
\begin{equation}\label{eq_wtd}
    \mathcal P(t) = \frac{\Tr{\left\{\Lambda_j \, e^{\mathcal L_{\rm{nh}}^j\,t}\,\Lambda_j\,\hat\rho\right\}}}{\gamma\Tr\left\{{\Lambda_j\hat\rho}\right\}}.
\end{equation}
Studying this object allows to unveil time correlations between jumps and provides an alternative framework to characterize phases of matter.

The aim of our work is to use the statistics arising from the objects introduced above to characterize the phase diagram in an alternative way. Given the many-body complexity of solving exactly the master equation generated by the tilted Lindbladian, or of computing the waiting-time distribution for the full chain, we specialize to a set of methods and approximations that allow us to perform calculations in a simplified setting while retaining the essential many-body features of the problem. In particular, we employ cluster mean-field theory, which enables us to capture any order of short-range correlations between spins, and a cumulant expansion, which instead accounts for long-range correlations up to a given order, both depicted in Fig.~\ref{fig:single-mf} (b) and (c). These two approaches provide complementary perspectives on the many-body nature of the model and, when combined, yield a more complete understanding of the system’s behavior.

\section{Full counting statistics of jumps}\label{sec_fcs}

In this section we solve the master equation generated by the $L_{\chi}$ by means of several approximations. We start from the easiest approximation given by mean-field, and then incorporate correlations between spins. In order to do so, we will focus first on the introduction of short-range correlations using cluster mean-field, and then on the insertion of long-range correlations through a cumulant expansion. The various methods and their assumptions will be introduced during the following discussion.

\subsection{Cluster mean-field approximation}

To perform a cluster mean-field approximation to the tilted Lindbladian dynamics, we start by assuming a quantum state which is factorizable in clusters of $N_c$ sites for the tilted density matrix,
\bea
    \rhohat_\chivec = \bigotimes_{\mu=1}^{N/N_c}\rhohat_{\chivec_\mu}^{(\mu)}\,,
\eea
similarly to the mean-field approximation of the untilted case in App.~\ref{app_dpt}.

Assuming translational invariance among the clusters, each cluster becomes equivalent to
the others, and one can focus on the state of one of them to reduce the complexity of the
problem. Furthermore, assuming that we are interested in counting only jumps on one cluster, we define two classes of density matrices -- one ($\mu=1$) where we count the dynamical activity, and that of all the other clusters where we do not.
In this sense, we set $\chivec_1=\chivec=(\chi_1,\cdots,\chi_{N_c})$ and $\chivec_{\mu\neq 1}=\Vec{0}$, which implies that 
\bea
    \rhohat_{\chivec_{\mu\neq 1}}^{(\mu)} = \rhohat^{(\mu)}_{\chivec_\mu=\Vec{0}}=\sum_{\nvec_\mu}\rme^{\rmi  \nvec_\mu\cdot \Vec{0}}\tilde\rhohat^{(\mu)}_{\nvec_\mu}=\rhohat^{(\mu)}\,,
\eea
namely the reduced states of the other clusters are normalized with unity trace.  We thus have
\begin{equation}\label{eq_ansatzTLmf}
    \oprho_{(\chivec,\Vec{0},\cdots,\Vec{0})}=\oprho^{(1)}_{\chivec}\bigotimes_{(\mu)\neq (1)}\oprho^{(2)}.
\end{equation}
Making these kinds of assumptions implies translational invariance for the unmonitored clusters despite the fact that the ones closest to the monitored one will feel a different dynamics because of the coupling with it. The latter is an approximation valid only for infinite- and long-range models in the thermodynamic limit. Indeed, for infinite-range models all degrees of freedom are coupled with equal strength, independently of their spatial separation. As a consequence, there is no underlying geometric structure that distinguishes clusters based on their position, and all unmonitored clusters are dynamically equivalent. In this case, the assumption of translational invariance among the unmonitored clusters is exact.

On the other hand, in long-range models the interaction strength depends on the distance between sites, thus introducing a nontrivial geometric structure. Nevertheless, in the thermodynamic limit the number of couplings involving non-monitored clusters only is extensive, while the couplings between the monitored cluster and the rest of the system give a subextensive contribution. As a result, deviations from translational invariance among the unmonitored clusters become negligible. Furthermore, the accuracy of this approximation improves as the cluster size increases. 

An alternative approach would be to consider a factorized Ansatz where each cluster has its own density matrix. Explicitly carrying out numerical calculations for finite sizes, one should be able to capture the classical contributions with higher accuracy, with the trade-off that analytical calculations become unfeasible. 

Plugging the new Ansatz into Eq.~\eqref{eq_tiltedL}, and tracing out all the degrees of freedom of the clusters $(\mu)\neq (1)$, we end up with two coupled cluster-mean-field equations, namely one for the tilted $\oprho_{\chivec}^{(1)}$ and another for the density matrix $\oprho^{(2)}$ that evolves under the standard mean-field Lindblad equation reported in App.~\ref{app_dpt}. The set of equations reads
\bea\label{eq_lind_cmf}
        \dfrac{\rmd\rhohat_\chivec^{(1)}}{\rmd t} &= -\rmi\[\Ham_{\scriptscriptstyle\mathrm{cMF}}\(\rhohat^{(2)}\),\rhohat_\chivec^{(1)}\]+\gamma\sum_{i=1}^{N_c}\dcal_{\chi_i}[\sigmam_i]\rhohat_\chivec^{(1)}\,,\\
        \dfrac{\rmd\rhohat^{(2)}}{\rmd t} &= -\rmi\[\Ham_{\scriptscriptstyle\mathrm{cMF}}\(\rhohat^{(2)}\),\rhohat^{(2)}\]+\gamma\sum_{i=1}^{N_c}\dcal[\sigmam_i]\rhohat^{(2)}\,,
\eea
where
$\dcal_{\chi_i}[\PauliSigma_i^-](\bigcdot)\equiv \rme^{\rmi\chi_i}\PauliSigma_i^-\bigcdot\PauliSigma_i^+-\dfrac{1}{2}\left\{ \PauliSigma_i^+\PauliSigma_i^-,\bigcdot \right\}$, and $\Ham_{\scriptscriptstyle{\mathrm{cMF}}}\(\rhohat^{(2)}\)$ depends on the standard mean-field values of $\PauliSigma^x$ on each site within the cluster:

\bea
 \Ham_{\scriptscriptstyle\mathrm{cMF}}\(\rhohat^{(2)}\) = \Ham^{\scriptscriptstyle\mathrm{int}}_{\scriptscriptstyle\mathrm{cMF}}\(\rhohat^{(2)}\)+h \sum_i^{N_c}\PauliSigma_i^z.
\eea
In turn, the first term in the r.h.s. reads
\begin{equation}
    \Ham^{\scriptscriptstyle\mathrm{int}}_{\scriptscriptstyle\mathrm{cMF}}\(\rhohat^{(2)}\)=\sum_{i=1}^{N_c}\Ham^{\scriptscriptstyle\mathrm{int},i}_{\scriptscriptstyle\mathrm{cMF}}-\frac{J}{\calN_\alpha}\sum_{i,j=1}^{N_c}\frac{1}{||i-j||^\alpha}\PauliSigma_i^x\PauliSigma_j^x\,.
\end{equation}
The first term in the summation, namely $\Ham^{\scriptscriptstyle\mathrm{int},i}_{\scriptscriptstyle\mathrm{cMF}}$, is the cluster-mean-field Hamiltonian for the $i$-th site within the cluster due to the interaction with the sites of the surrounding clusters, while the second term represents the interaction between spins of the same cluster. The former is the standard cluster mean field (see App.~\ref{app_dpt} for more details), which is made up of single-site terms of the form:
\begin{equation}
\begin{split}
    \Ham_{\scriptscriptstyle\mathrm{int}}^{\scriptscriptstyle\mathrm{cMF},i}=-2\frac{J}{\calN_\alpha}\PauliSigma_i^x\sum_{j=1}^{N_c}
    &m_j^x N_c^{-\alpha}[\zeta(\alpha,(i-j)/N_c+1)\\
    &+\zeta(\alpha,(j-i)/N_c+1)],
\end{split}
\end{equation}
where $\zeta(\alpha,q)=\sum_{\xi=0}^{\infty}\sfrac{1}{(\xi+q)^\alpha}$ is the Hurwitz Zeta function, deriving from the notion of distance in PBC (see App.~\ref{app_dpt} for more details), and $m^x_j = \Tr\sigmax_j\rhohat^{(2)}$.

\begin{figure}[ht]
    \centering
    \includegraphics[width=\columnwidth]{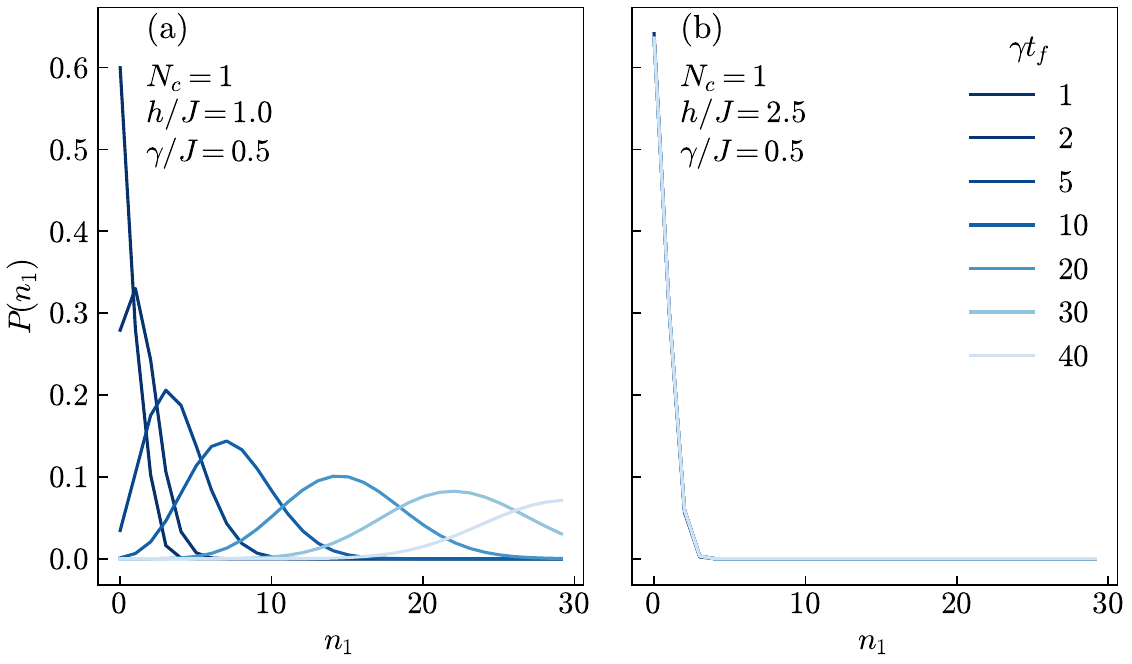}
    \caption{Evolution of $P(n_1)$ in the tilted mean-field approximation with $N_c=1$ in ferromagnetic (a) and paramagnetic (b) phases. $\alpha=1.1$.}
    \label{fig:pnt-single-mf}
\end{figure}

In the following, we will always assume $\alpha = 1.1$.
For $N_c=1$, the mean-field Hamiltonian reduces to
\begin{equation}
    \hat H_{\scriptscriptstyle{1\mathrm{MF}}} = h\hat\sigma^z -2Jm_x\hat\sigma^x.\label{eq:hmf}
\end{equation}
Solving the tilted mean-field equations for a single-site cluster ($N_c=1$), we analyze the probability distribution $P(n_1,t)$ of the total number of quantum jumps occurring on the monitored site within growing time intervals $[0,t_f]$.
In the ferromagnetic phase ($h=1.0$), shown in Fig.~\ref{fig:pnt-single-mf}(a), the distribution progressively broadens and shifts towards larger values of $n_1$ as
$t_f$ increases. This reflects the persistent activity of the dissipative dynamics, which leads to a growing number of jump events over time.
In contrast, in the paramagnetic phase ($h=2.5$), Fig.~\ref{fig:pnt-single-mf}(b), the distribution remains sharply peaked at small values of $n_1$ even for large $t_f$. Within the single-site mean-field approximation, this behavior originates from the emergence of a dark state of the Lindblad dynamics, for which the jump rate vanishes and the dissipative evolution effectively freezes at long times. 

This complete suppression of jump activity is, however, an artifact of the cluster mean-field approximation with $N_c=1$, which neglects spatial correlations and dynamical fluctuations. In more refined approaches that include correlations beyond mean field, such as the cumulant expansion discussed below, we expect the dark state to no longer be strictly absorbing and the residual jump activity to restore a more faithful description of the dissipative phase transition.

When $N_c>1$, we can retrieve the probability distribution of having $\nvec= (n_1,n_2,\cdots, n_{N_c})$ jumps on sites $1,\cdots, N_c$. 

\begin{figure}[ht]
    \centering
    \includegraphics[width=\columnwidth]{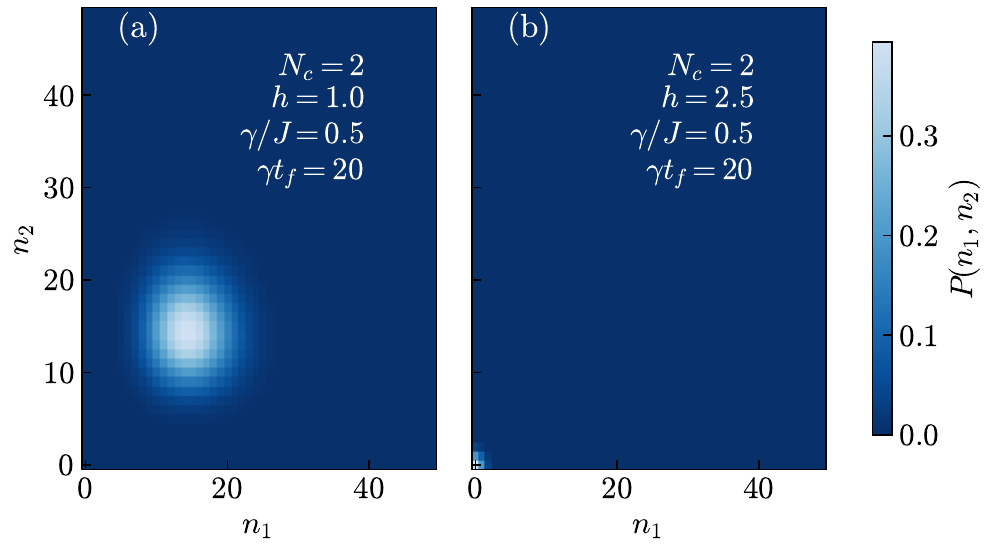}
    \caption{Probability distribution $P(n_1,n_2)$ for $\gamma t_f=20$, $N_c=2$, $\gamma/J = 0.5$, $\alpha=1.1$. (a) Ferromagnetic phase. (b) Paramagnetic phase.}
    \label{fig:heatmap_pn1n2_nc_2}
\end{figure}

As for the $N_c=1$ case, we can see that, as time grows, two distinct behaviors characterize the ferromagnetic and the paramagnetic phases, which emerges in Fig.~\ref{fig:heatmap_pn1n2_nc_2}. In panel (a), representing the ferromagnetic regime at $\gamma t_f=20$ the light circular shape represents a peak in the probability distribution $P(n_1,n_2)$ around the values $n_1,n_2\simeq 15$. In this phase, the peak moves towards higher values of $n_1$ and $n_2$ with time and the joint distribution progressively broadens.  On the other hand, in panel (b), representing the paramagnetic regime, we can again see the artifact of the mean-field approximation manifesting with a $P(n_1,n_2)$ that is extremely peaked around small values of $n_1$ and $n_2$.

\begin{figure}[ht]
    \centering
    \includegraphics[width=\columnwidth]{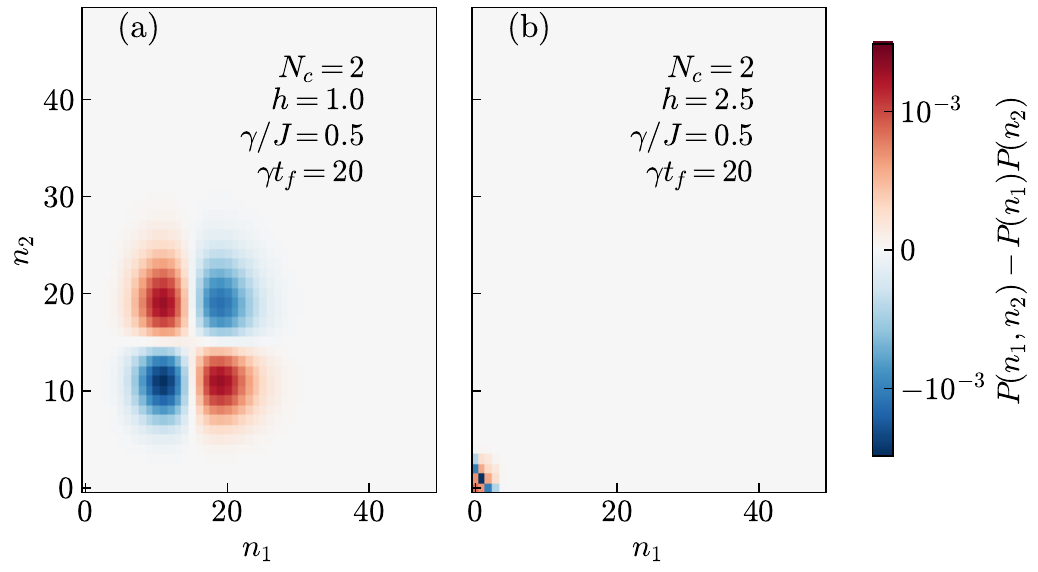}
    \caption{Connected joint distribution $P(n_1,n_2)-P(n_1)P(n_2)$ with $N_c=2$, $\gamma/J = 0.5$, $\gamma t_f=20$, and (a) $h=1.0$ (b) $h=2.5$. $\alpha=1.1$.}
    \label{fig:heatmap_connected_pn1n2_nc_2_vs_time}
\end{figure}

To further probe the effects of short-range correlations among the two-site cluster, considering the marginals $P(n_1)=\sum_{n_2}P(n_1,n_2)$ and $P(n_2)=\sum_{n_1}P(n_1,n_2)$, we study the connected joint distribution $P(n_1, n_2)-P(n_1)P(n_2)$, reported in Fig.~\ref{fig:heatmap_connected_pn1n2_nc_2_vs_time}. This quantity is useful, as it signals correlated or anti-correlated jumps whenever it is different from zero.
The latter reveals the emergence of non-trivial dynamical correlations in the ferromagnetic phase. Indeed, in this regime (panel (a)), we observe a four-quadrant structure with positive weight in the mixed sectors ($n_1>\mathbb{E}[n_1]$, $n_2<\mathbb{E}[n_2]$ and vice versa). This is an indication of anti-correlated fluctuations around the mean number of jumps. On the other hand, in the paramagnetic regime (panel (b)), the distribution is factorized. 

\begin{figure}[ht]
    \centering
    \includegraphics[width=\columnwidth]{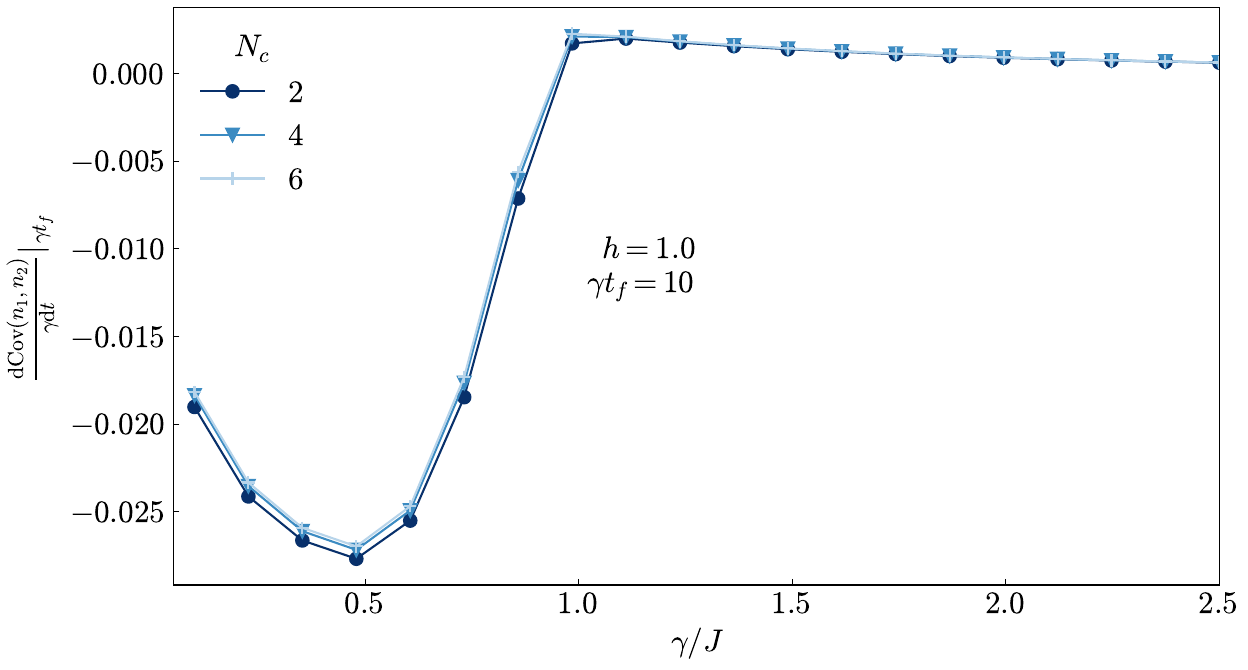}
    \caption{Growth rate of the covariance of the number of jumps in time from the cluster mean field approximation with $N_c=2,4,6$, $h = 1.0$, $\gamma t_f=10$, $\alpha=1.1$.}
    \label{fig:covariance_cmf}
\end{figure}

These features can be further found by looking at the cross-cumulants of the joint probability distribution of the dynamical activity. In Fig.~\ref{fig:covariance_cmf} we analyze the covariance

\begin{equation}
\begin{split}
    &\text{Cov}(n_1,n_2)=\sum_{n_1,n_2\in\calN}n_1 n_2 P(n_1,n_2)\\
    &-\sum_{n_1\in\calN}n_1 P(n_1)\sum_{n_2\in\calN}n_2 P(n_2),
\end{split}
\end{equation}
where $\calN$ is the range of values that $n_i$ can reach during the dynamics on each site $i$, and the time dependence is left implicit. We refer to sites $1$ and $2$ to indicate the central sites of the cluster regardless of its size, namely $1\equiv \sfrac{N_c}{2}-1$, $2\equiv \sfrac{N_c}{2}$.
Since in the ferromagnetic phase the number of jumps increases linearly with time during the dynamics, we know that both $\mathbb{E}[n_1,n_2]$ and $\mathbb{E}[n_1]\mathbb{E}[n_2]$ will grow quadratically. When subtracting one from the other, the leading quadratic contribution cancels out and a subleading linear contribution persists, so that the covariance decreases linearly with time. For this reason, we analyze its time-scaled analogs, namely the growth rate of the covariance in the stationary regime $\sfrac{\text{d}\text{Cov}(n_1,n_2)}{\gamma\text{d}t}(t)$. In the following, we show the results for growing values of $N_c$ in Fig.~\ref{fig:covariance_cmf}. 
In it, the time derivative appears to be different from zero and negative in the range of values corresponding to the ferromagnetic phase, while going to zero in the paramagnetic phase, again signaling a qualitative difference between the two dissipative phases.

\subsection{Cumulant expansion} 
To go beyond the mean-field approximation and include long-range correlations in the study of counting statistics, we propose in this section the cumulant expansion approximation for solving the tilted Lindbladian evolution. To proceed, let us consider the tilted Lindbladian dynamics in its adjoint form. Defining the trace-normalized tilted density matrix,
\bea
    \rhotil_\chivec \equiv \dfrac{\rhohat_\chivec}{\cscr_\chivec}\,,\quad \cscr_\chivec\equiv \Tr[\rhohat_\chivec]\,,
\eea
it can be interpreted as an analogy of a distribution, and we denote the ``tilted expectation'' of an observable $\Ohat$ as $\langle\Ohat\rangle_\chivec \equiv \Tr[\rhotil_\chivec\Ohat]$. The trace $\cscr_\chivec$ evolves as 
\bea\label{eq:der-trace}
    \dfrac{\rmd}{\rmd t}\cscr_\chivec = \cscr_\chivec\sum_j(\rme^{\rmi\chi_j}-1)\langle\dLhat_j\Lhat_j\rangle_\chivec\,,
\eea
from which we obtain the dynamics of $\rhotil_\chivec$ in the form of $\rmd\rhotil_\chivec/\rmd t = \tilde\Lcal_\chivec\rhotil_\chivec$, with the trace-normalized tilted Lindbladian defined as,
\bea
    \tilde\Lcal_\chivec  &\equiv 
     \Lcal_\chivec - \sum_j(\rme^{\rmi\chi_j}-1)\langle\dLhat_j\Lhat_j\rangle_\chivec\,.
\eea
This allows us to define the adjoint normalized tilted Lindbladian via $\Tr[\tilde\Lcal_\chivec(\rhotil_\chivec)\Ohat]\equiv\Tr[\rhotil_\chivec\tilde\Lcal_\chivec^\ddagger(\Ohat)]$, for a time-independent operator $\Ohat$,
\bea
    \tilde\Lcal_\chivec^\ddagger\Ohat &= \rmi [\Ham,\Ohat] \\&+ \sum_j\left( \rme^{\rmi\chi_j}\dLhat_j\Ohat\Lhat_j - \dfrac{1}{2}\left\{\dLhat_j\Lhat_j,\Ohat\right\} \right)\\
    &-\sum_j(\rme^{\rmi\chi_j}-1)\langle\dLhat_j\Lhat_j\rangle_\chivec\Ohat\,,
\eea
providing a convenient adjoint picture for the operator's evolution under the tilted Lindbladian via the relation $\rmd\langle\Ohat\rangle_\chivec/\rmd t = \langle \tilde\Lcal_\chivec^\ddagger\Ohat\rangle_\chivec$.
As in the standard Lindbladian case, this typically leads to an infinite hierarchy of equations involving nonlocal moments $\langle\prod_i\sigmahat_i^{\alpha_i}\rangle$ of increasing orders. An approximation can be made by performing a truncation, such that all cumulants higher than a certain order $k_c$ are assumed to be zero. (Note that since the single-site Pauli operators form a closed multiplication algebra, products of operators acting on the same site can be exactly evaluated.) The case $k_c=1$ then corresponds to the standard single-site mean-field approximation obtained with a Gutzwiller product-state ansatz. In the following, we will focus on 
the case $k_c=2$, allowing for beyond-mean-field long-range correlations to be captured on the level of quadratic fluctuations.

\begin{figure*}
    \includegraphics[width=\linewidth]{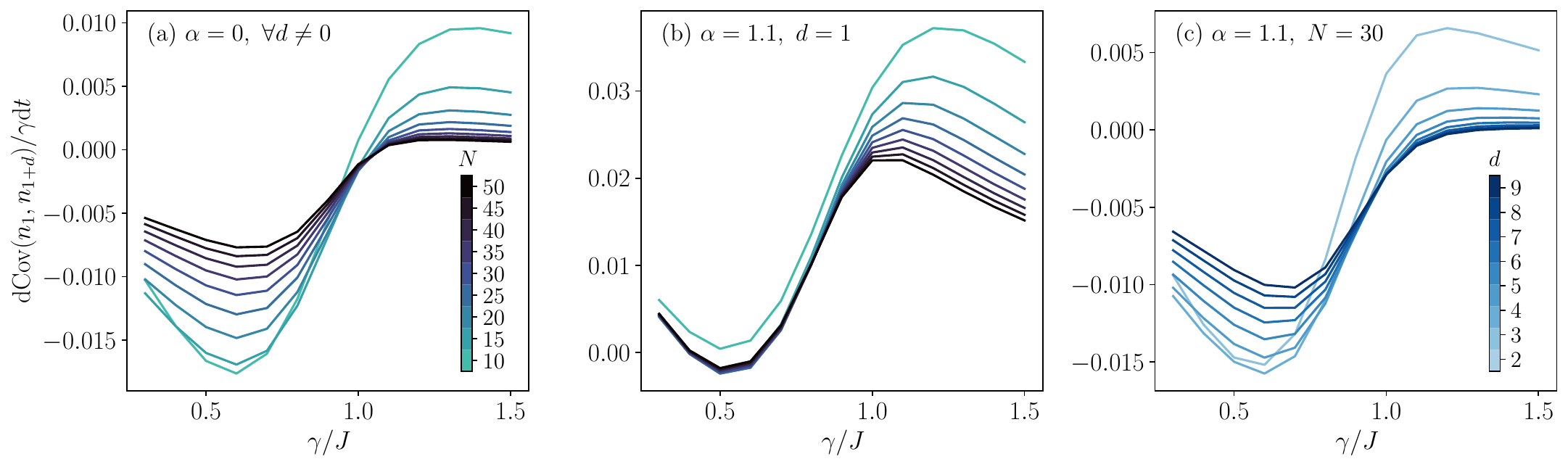}
    \caption{Steady-state rate of change of the covariance between the numbers of jumps on sites $1$ and $1+d$ as a function of the dissipation $\gamma$ on the periodic spin chain with $h=J$, obtained with the cumulant expansion approach to the tilted Lindblad equation. (a) Results for $\alpha=0$ and different system sizes $N$ (see legend). Due to the infinite-range interaction, the two-site jump correlation is independent of the distance $d$. (b) Results for $\alpha=1.1$ and nearest-neighbor sites ($d=1$), for different system sizes $N$ [legend shared with panel (a)]. (c) Results for $\alpha=1.1$ and different distances $d$ (see legend), for a fixed system size of $N=30$.}
    \label{fig:cum_tilt_cov}
\end{figure*}

In this second-order cumulant truncation scheme, the state can be fully characterized by the first-order cumulants $c^\alpha_j\equiv\langle\sigmahat_j^\alpha\rangle_\chivec$ and the second-order ones, $v^{\alpha\beta}_{jk}\equiv \langle\sigmahat_j^\alpha\sigmahat_k^\beta\rangle_\chivec-c_j^\alpha c_k^\beta = \langle\delhat_j^\alpha\delhat_k^\beta\rangle_\chivec$, with $\delhat_j^\alpha\equiv \sigmahat_j^\alpha-\langle\sigmahat_j^\alpha\rangle_\chivec$. The complete set of equations of motion can then be formally derived as (See App.~\ref{app:cum}),
\bea\label{eq:cum-tilt}
    \dfrac{\rmd}{\rmd t}c^\alpha_j &= \langle\tilde\Lcal^\ddagger_\chivec\sigmahat_j^\alpha\rangle_\chivec\,,\\
    \dfrac{\rmd}{\rmd t}v^{\alpha\beta}_{kl}&= \langle\tilde\Lcal^\ddagger_\chivec(\delhat_k^\alpha\delhat_l^\beta)\rangle_\chivec\,,
\eea
which, combined with Eq.~\eqref{eq:der-trace} for the trace $\cscr_\chivec$, determines the counting statistics via the Fourier relation~\eqref{eq:pn-fourier}. A few remarks are in order.
\begin{itemize}
    \item In the equations above, all terms beyond order $k_c=2$ should be decoupled according to the truncation rule that cumulants of order higher than $k_c$ are enforced to be zero.
    \item The standard Lindblad case can be trivially recovered by setting $\chivec=\Vec{0}$ in the equations.
    \item As the second-order cumulant truncation for spin operators does not correspond to a variational ansatz for the state, the approximation is not controlled and may lead to unphysical results~\cite{Verstraelen2023}.
\end{itemize}

Contrary to the cluster mean-field approximation, which truncates any correlation beyond the cluster size $N_c$, the cumulant approach includes long-range correlations (up to the quadratic level), which is suited for probing long-range interacting systems and better captures finite-size effects.

As an illustration, let us consider the jump correlation between two sites separated by a distance $d$ on the periodic spin chain, quantified by the rate of change of the covariance, $\rmd\mathrm{Cov}(n_1,n_{1+d})/\gamma\rmd t$, which admits a steady-state value at long times as a consequence of the central limit theorem. Fig.~\ref{fig:cum_tilt_cov} (a) shows the result for an infinite-range interacting system ($\alpha=0$), where the dependence on the distance $d$ is absent. 
Due to the mean-field nature of the setup, we recover qualitatively similar phenomenology to the cluster mean-field prediction shown in Fig.~\ref{fig:covariance_cmf}. In the ordered phase ($\gamma<\gamma_c\simeq J$), the dynamics is dominated by the interaction $J$ term, driving the system into a state with long-range coherence where the excitations are highly nonlocal. 
A local jump event, therefore, temporarily depletes the collective excitation and suppresses neighboring jumps, resulting in anti-correlation in $P(n_i,n_j)$, signaling the spatial antibunching of jumps. 
In the disordered phase, the high dissipation $\gamma$ destroys the long-range correlations, and the infinite-range interaction merely serves as a background mean field driving the otherwise independent sites.  A local jump in this regime then signals an ephemeral excitation of the global mean field, resulting in slightly bunched jumps, as evidenced by a positive correlation, which is eventually washed out in the large-system-size limit, where the sites become decoupled. The correlation crosses zero around the critical point due to the competition between the coherent interaction and the dissipation. 

Fig.~\ref{fig:cum_tilt_cov} (b) shows the results for nearest-neighbor jump correlations ($d=1$) with a shorter interaction range $\alpha=1.1$, where the above mean-field picture breaks down. Despite the qualitative similarity to the all-to-all interacting case, the spatially localized correlator $\mathrm{Cov}(n_1,n_2)$ is enhanced by the increased fluctuations close to criticality. This leads to a peak in the jump correlations close to the phase transition, which becomes more prominent with increasing system sizes. Fig.~\ref{fig:cum_tilt_cov} (c) shows the results for a fixed system size $N=30$ with the same interaction range $\alpha=1.1$, yet for different distances $d$ between the two probed sites. At larger distances, the contribution from local fluctuations fades away, and the correlations are again dominated by the long-wavelength mean-field-like excitations, and we recover similar phenomenology to the infinite-range scenario displayed in panel (a).

\section{Waiting-time distribution}\label{sec_wtd}
This section focuses on the statistics of time intervals occurring between two jumps. As in the previous Section, the study of this quantity will first be performed in the simpler setting of mean-field, and then correlations between jumps will be added through a cluster mean-field approximation of the setting.

To overcome the complexity of calculating the full many-body waiting-time distribution, we will rely on a specific setting in which, through cluster mean-field we can get good analytical and numerical predictions.

First of all, we suppose to monitor one site only of the whole chain. In our case we will always focus on the first site. All the other $N-1$ sites evolve according to the Lindblad equation and the only readout information we keep for statistics regards the jumps happening on the first site.
In a cluster mean-field perspective, we assume the following Ansatz for our quantum state. For clusters of size $N_C$:
\begin{equation}
    \hat \rho = \hat R^{(0)}\bigotimes_{\mu=1}^{N/N_c} \hat R^{(\mu)} = \hat R^{(0)}\bigotimes_{\mu=1}^{N/N_c} \hat R^{\rm{cMF}} .
\end{equation}
The first cluster $\mathcal C_0$, which contains the monitored spin, is allowed to have a different state with respect to all the other un-monitored clusters $\mathcal C_{\mu>1}$, which are instead assumed to be all the same.
Specifically, the dynamics of the $N/N_C-1$ non-monitored clusters is then assumed to be generated by the Lindblad equation~\eqref{eq_lind_cmf}. The cluster with the monitored spin evolves instead according to the Hamiltonian
\begin{equation}\label{eq_wtd_cmf}
    \hat H_0^{\rm{{cMF}}} = -\frac{J}{\mathcal N_\alpha}\sum_{i,j\in \mathcal C_0}\frac{\hat\sigma_i^x\hat\sigma_j^x}{r_{ij}^\alpha}-2J\sum_{i\in\mathcal C_0}g_i\hat\sigma_i^x+h\sum_{i\in\mathcal C_0}\hat\sigma_i^z,
\end{equation}
and quantum jump measurements. Notice that the Hamiltonian contains the parameter
\begin{equation}
    g_i = \sum_{j\not\in\mathcal C_0}\frac{\expval{\hat\sigma_j^x}}{\mathcal N_\alpha\, r_{ij}^\alpha},
\end{equation}
which contains information about the Lindblad evolution of the non-monitored clusters.

The Ansatz and the dynamics we are proposing in this scenario are quite similar to what we proposed for the tilted Lindbladian case: while the first cluster is subject to monitoring (or counting, in the case of the full counting statistics), all the other clusters will evolve without keeping track of measurement records and are assumed to be translationally invariant. As already discussed for the tilted Lindbladian case, this Ansatz yields exact results in the thermodynamic limit and the long-range regime. In these cases, the interaction terms between the monitored cluster and its neighbouring Lindblad clusters are subextensive and can be neglected. For shorter ranges, $\alpha\gtrsim1$, our Ansatz provides anyway a good approximation for the bulk of the Lindblad clusters.

We now first focus on the single-site mean field. This does not account for correlations between spins, and allows us to consider the first cluster composed only of the spin we are monitoring. This procedure yields a simple description in terms of the monitored spin only whose dynamics is governed by the simple Hamiltonian in Eq.~\eqref{eq:hmf}
and quantum jumps. Notice that in this case we have $g_i=m_x=\expval{\hat\sigma_i^x}$ $ \forall i>1$.

Dealing with a single-site problem, we can use Eq.\eqref{eq_wtd} directly to calculate the waiting time distribution. In this case, the non-Hermitian Hamiltonian responsible for the smooth evolution between jumps is $\hat H_{\rm{nh}}=\hat H_0^{\rm{MF}}-\gamma/2\,\hat\sigma_1^+\hat\sigma_1^-$. The mean-field Hamiltonian for the first site has as input parameter $m_x$, given by the Lindblad evolution of the other spins: in order to calculate the WTD in the steady-state, we set $m_x$ to its steady state value:
\begin{equation}
    m_x^* = \begin{cases}
            \sqrt{\frac{-h^2+2hJ-\gamma^2}{2J^2}} & \text{ if }-h^2+2Jh-\gamma^2>0, \\
            0 & \text{ elsewhere },
            \end{cases}
\end{equation}
This dramatically simplifies the structure of the waiting-time distribution in Eq.~\eqref{eq_wtd}
\begin{equation}
    \mathcal P(t) = \frac{\Tr{\left\{\hat\sigma_1^-\,\,e^{-it\hat H_{\rm{nh}}}\,\hat\sigma_1^-\,\hat R_0\,\hat\sigma_1^+\,e^{it\hat H_{\rm{nh}}^*}\,\hat\sigma_1^+\right\}}}{\Tr{\left\{\hat\sigma_1^+\hat\sigma_1^-\hat R_0\right\}}},
\end{equation}
and allows for an analytical calculation of the full structure of the waiting time distribution, which yields

\begin{equation}\label{eq_wtdMF}
    \mathcal P(t)=\frac{16e^{-2\gamma t}J^2m_x^{*^2}\gamma\,\sin\left(t\Gamma\right)\,\sin\left(t\Gamma^*\right)}{|\Gamma|^2}
\end{equation}
with $\Gamma = \sqrt{4J^2m_x^{*^2}+(h-i\gamma)^2}$.
Notice how $m_x^*$ multiplies the whole expression. In the paramagnetic region, where $m_x^*=0$, the WTD tends to the uniform distribution over the interval $t\in[0,\infty]$. 
Additionally, note that there is no dependence on the initial state of the first site $\hat R_0$ from Eq.\eqref{eq_wtd} as the initial jump always projects it into $\ket{\downarrow}\bra{\downarrow}$.

Eq.~\eqref{eq_wtdMF} allows to calculate exactly the average waiting time and its variance, which both can be used as order parameters to describe statistical properties of the dissipative phases of our model.
In particular, the average waiting time reads
\begin{equation}
    \mathbb E_{\scriptscriptstyle\mathrm{MF}}[t]=\int dt\, \mathcal P(t)\, t = \frac{h^2+2J^2{m_x^*}^2+\gamma^2}{4J^2{m_x^*}^2\gamma},
\end{equation}
while the variance reads

\begin{align}
    &\mathrm{Var}_{\scriptscriptstyle\mathrm{MF}}[t] = \int dt\, \mathcal P(t)\, t^2\,-\,\mathbb E_{\scriptscriptstyle\mathrm{MF}}[t]^2 =\\& \frac{h^4+6h^2J^2{m_x^*}^2+4J^4{m_x^*}^4+2\gamma^2(h^2-J^2{m_x^*}^2)+\gamma^4}{16J^4{m_x^*}^4\gamma^2}. \nonumber
\end{align}
Since both the average and the variance have the denominator multiplied by $m_x^*$, we notice that how the distribution approaches the uniform one yields an infinite average and variance.
Using these facts, we can characterize the ferromagnetic phase as the region with finite average waiting time and variance, and the paramagnetic phase as the region with infinite average waiting time and variance. Notice that this is consistent with the mean-field predicting the approach to a dark state in the paramagnetic region: when no jump occurs, the average waiting time diverges. 

It is interesting now to add back correlations between spins in this description, and consider the case of larger clusters containing more than one site. In this case we are no longer able to give a single-site description of the monitored spin, but we should describe carefully the dynamics of the whole cluster containing the monitored spin, and do not keep track of the measurement outcomes of the non-monitored spins inside the same cluster. 

Despite looking like a complex task, the cluster mean-field approximation allows for favourable numerical simulations, as one only needs to simulate the dynamics of $N_c$ spins instead of the whole chain.
On a practical level, we first run the cluster mean field simulations for the Lindblad dynamics of the $N/N_c-1$ non-monitored clusters. This produces the external field $g_i$, $i=1,..,.N_C$ which can be used to evolve the first cluster containing the monitored spin.
We then use $\hat H_0^{\rm{cMF}}$ from Eq.~\eqref{eq_wtd_cmf} to simulate the dynamics of the cluster, and collect data from the monitoring of the site $j=1$ through a Monte Carlo algorithm. We don't keep information about the monitoring of the other sites inside the first cluster, so that they effectively evolve with Lindblad.

\begin{figure}
    \centering
    \includegraphics[width=1.1\linewidth]{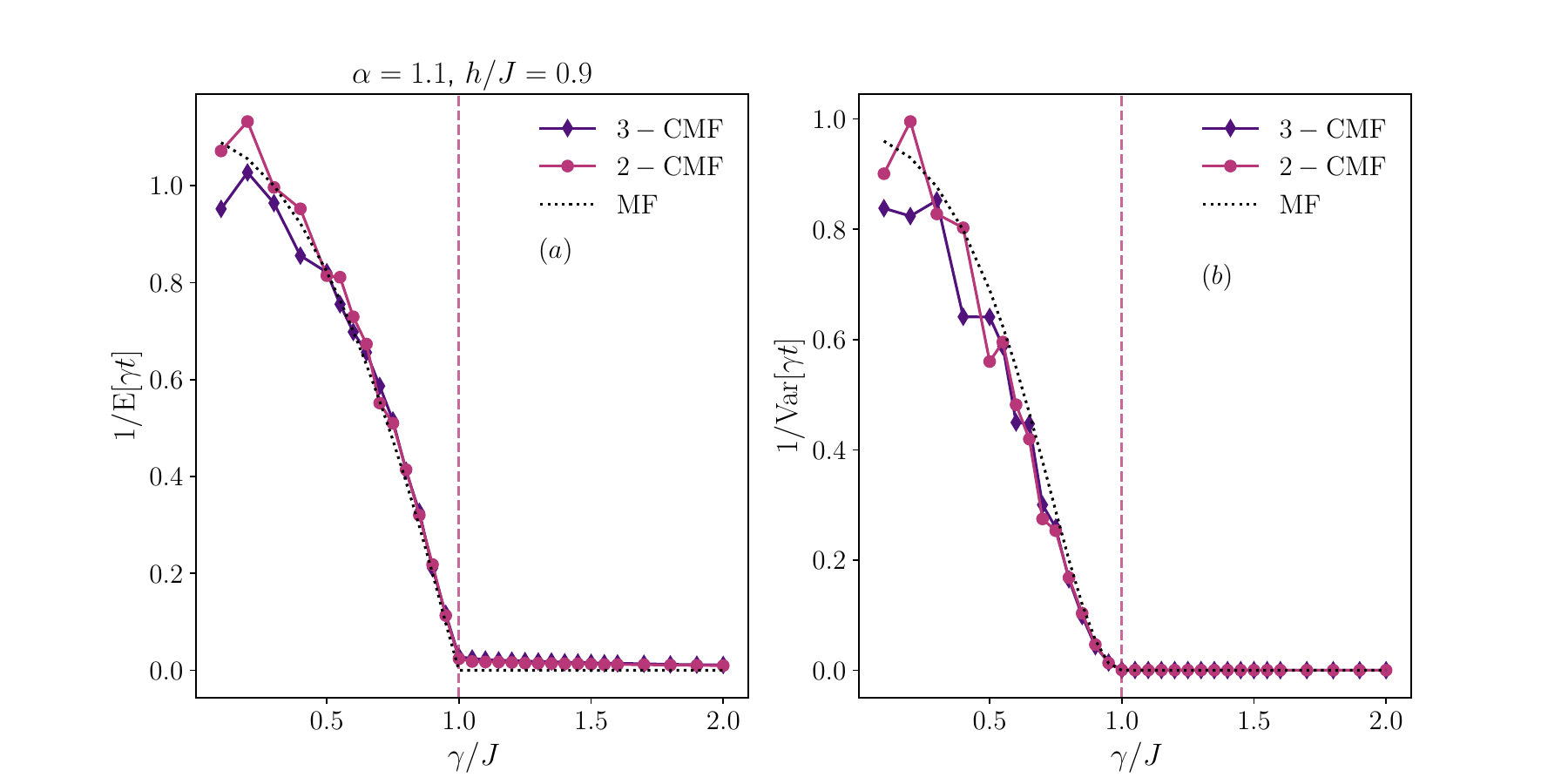}
    \caption{Inverse of (a) average  (b) variance of waiting time distribution at $\alpha=1.1<\alpha_C$. The dotted line represents mean-field, the vertical dashed lines represent the separation between the two regions predicted by cluster mean-field. Cluster mean-field converges and confirms the presence of two distinct regions.}
    \label{fig:cmfwtd1.1}
\end{figure}

\begin{figure}
    \centering
    \includegraphics[width=1.1\linewidth]{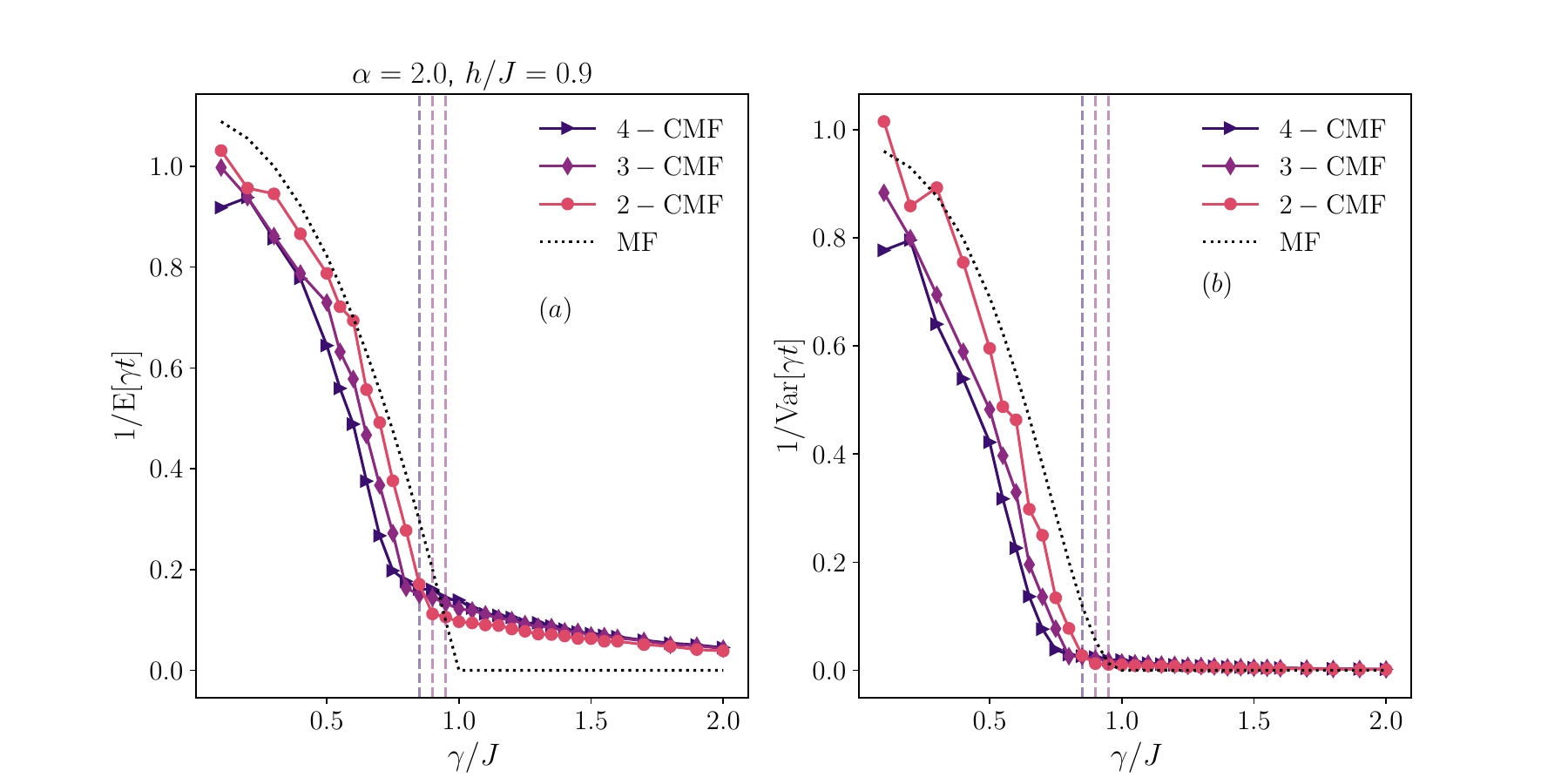}
    \caption{Inverse of (a) average  (b) variance of waiting time distribution at $\alpha=2.0 \simeq \alpha_C$. The dotted line represents the mean-field, the vertical dashed lines represent the separation between the two regions predicted by cluster mean-field. Cluster mean-field shrinks more and more the area corresponding to finite variance.}
    \label{fig:cmfwtd2.0}
\end{figure}

Through these methods, we are able to obtain numerically an estimate for the average and the variance of the waiting time,
which are suggestive of the dissipative phase diagram,
in great agreement with the predictions made on the Lindblad level via cluster mean field. 
For $\alpha=1.1$, the dissipative phase diagram presents at $h/J=0.9$ a phase transition between a ferromagnetic region to a paramagnetic region at $\gamma/J\sim1$. Both the inverse of the waiting time in Fig.~\ref{fig:cmfwtd1.1}  in panel (a), and the variance of the waiting time in panel (b), show similar features:
they are both zero for $\gamma/J\gtrsim1$ and finite for $\gamma/J\lesssim1$, according to $2$- and $3-$ cluster mean field, which show convergence around the transition point.
Once again, jump statistics features are able to reproduce the dissipative phase diagram: the ferromagnetic region is characterized by finite average and variance of the waiting time, while the paramagnetic region is characterized by large tails of the WTD and infinite average and variance.
It is interesting to note how both $2-$ and $3-$ cluster mean field predictions for the average and variance of waiting time distribution deviate slightly from the mean-field prediction (dotted line): short-range correlations taken into account by the cMF calculation are responsible for this small deviation.

Deviations from mean-field become more and more important as the range of interactions decreases. The case of $\alpha=2\simeq \alpha_C$ is presented in Fig.~\ref{fig:cmfwtd2.0}, and our approximation predicts the further shrinkage of the region with finite average and variance of waiting time. Panel (a) shows the inverse of the average waiting times: increasing the size of the cluster reduces the region with the biggest values of $1/\mathbb E[\gamma t]$ and increases the tail for $\gamma/J>1$: short-range interactions within the cluster allow for more frequent jumps on the site we are monitoring. A similar behaviour is observed in the inverse of the variance. Both quantities suggest a match with the predictions made at the level of the Lindblad equation, which, through cluster mean field, suggest the eventual disappearance of the ferromagnetic phase for $\alpha>\alpha_C$.

\section{Conclusions}
\label{sec_conclusions}
In this work, we have shown that the statistics of quantum jumps provide a direct trajectory-based characterization of dissipative phase transitions in a long-range open spin system.
By combining a tilted Lindbladian formalism with cluster mean-field approximation and a cumulant expansion, we have analyzed both the full counting statistics of the dynamical activity and the waiting-time distribution between consecutive quantum jumps. 

Within the cluster mean-field approximation, we found that the ferromagnetic phase is characterized by a persistent and anti-correlated jump activity, through the analysis of the joint distribution of jumps, their connected part, and the time derivative of their covariance.  In the paramagnetic phase, instead, the same quantities reveal that the jump statistics become weaker and approximately factorized. In this context, the waiting-time distribution provides a complementary indication of the dissipative phase transition, as analytically seen in a single-site mean field approach where its mean and 
variance are finite in the ferromagnetic regime, while diverging in the paramagnetic one. Considering larger clusters preserves this qualitative distinction between the two phases.

To complement the short-range picture, we developed a second-order cumulant expansion for the tilted Lindblad equation. This approach captures long-range correlations and finite-size effects, showing that in the infinite-range limit the correlations of the jumps are independent of the distance and change sign across the transitions, while, for power-law interactions, nearest-neighbor correlations are strongly enhanced close to criticality and become more mean-field like at larger distances.

Overall, the obtained results show that the cluster mean-field approximation and the second-order cumulant expansion provide complementary perspectives on the characterization of the dissipative phase transitions and on the related jump statistics, and highlight the potential of full counting statistics and waiting-time distribution of quantum jumps as probes of collective behavior in open many-body quantum systems beyond conventional steady-state observables.

Our work opens the way to several perspectives. The same methods could be applied to the study of non-trivial phases of matter, such as time-ordered phases like time crystals~\cite{wilczek2012quantum,khemani2019brief,sacha2020time,iemini2018boundary}, as the space-time correlations of quantum jumps could offer a different perspective on the characterization of these phases~\cite{viotti2026quantum}. One can also adopt complementary approaches to access the counting statistics via direct simulation of quantum trajectories within the semiclassical approximations~\cite{verstraelenGaussianQuantumTrajectories2018,liGeneralizedStochasticSpinwave2026}. To explore more exotic phases of matter, beyond-semiclassical methods~\cite{toscaEfficientVariationalDynamics2025,toscaVariationalDynamicsOpen2026,Maki2023} could be applied to solve quantum jump statistics and to characterize the tilted Lindbladian and the waiting time distribution, providing a fully quantum description of their behavior.

Moreover, given the direct connection to quantum measurement, it would be interesting to study the connection between quantum jumps behavior and their full counting statistics with measurement-induced phase transitions~\cite{yamamoto2026_01,tirrito2023full}, which have played a central role in recent studies of monitored quantum many-body systems~\cite{Turkeshi2021,skinner2019measurement,li2019measurement,cao2019entanglement}.

\textit{Acknowledgements} -- 
We thank Juan P. Garrahan and Igor Lesanovsky for useful feedback on the manuscript.
This work has been supported by the
European Union (ERC, RAVE, Grant No. 101053159); AD acknowledges funding from the European Research Council (ERC) under the European Union’s Horizon 2020 research and innovation programme (Grant agreement No. 101002955 — CONQUER); by the European Union -- NextGeneration EU within PRIN 2022, PNRR M4C2, Project TANQU 2022FLSPAJ [CUP B53D23005130006]; by the Provincia Autonoma di Trento;
\appendix

\begin{figure*}
    \centering
    \includegraphics[width=0.97\textwidth]{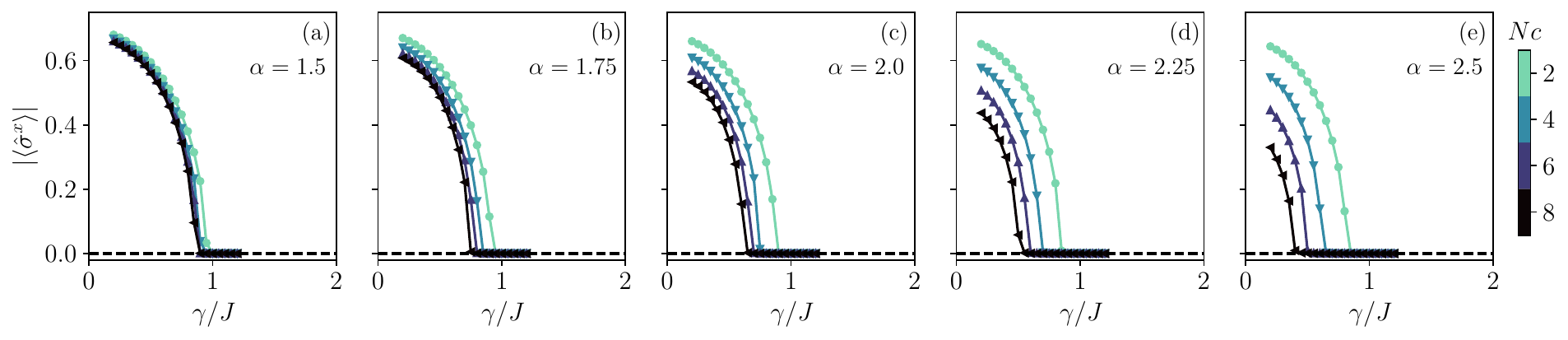}
    
    \caption{Cluster mean field results for the steady-state magnetization $|\langle\sigmahat^x\rangle|$ as a function of the dissipation rate $\gamma$ at $h=J$ for different cluster size $N_c$ and interaction range $\alpha$ (see legend).}
    \label{fig:cluster-mf-compare}
\end{figure*}

\section{Cluster mean field approximation}\label{app_dpt}
In this appendix, we derive the cluster mean field equations. We consider a cluster of sites as evolving under the effect of a mean field generated by the surrounding clusters, together with the effect of dissipation~\cite{sierant2022dissipative, fazio2025manybodyopenquantumsystems}. 
Within this approximation, one can assume
\bea\label{eq_ansatzMF}
    \rhohat = \bigotimes_\mu^{N/N_c} \rhohat^{(\mu)}\,,
\eea
where $\rhohat^{(\mu)}$ represents the density matrix for the $(\mu)$-th cluster of $N_c$ sites. Upon imposing translational invariance among the clusters, one can focus on the state of the first cluster, namely $\rhohat^{(1)}=\rhohat^{\scriptscriptstyle\mathrm{cMF}}$, without loss of generality.

We denote $\xi=0,...,\sfrac{N}{2N_c}$ (assuming $N$ is a multiple of $2N_c$) the index for the clusters, and $i,j=1,...,N_c$  the index for the sites within the cluster of the chain with PBCs. 
One can define $d^\rightarrow_\xi(i,j)=\xi N_c+j-i$ as the distance between the sites $i$ and $j$ in two respective clusters (separated by $\xi$ clusters), assuming that they are separated by fewer than $\sfrac{N}{2}$ sites when going right from $i$ to $j$. Analogously, one can set $d^\leftarrow_\xi(i,j)=\xi N_c+i-j$ as the same distance when the two sites $i$ and $j$ are separated by more than $\sfrac{N}{2}$ sites when going to the right, which, under PBCs, is equivalent to moving left.

In this setting, tracing out the degrees of freedom of all but the first cluster, the single-cluster Hamiltonian reads
\bea
 \Ham_{\scriptscriptstyle\mathrm{cMF}}(\rhohat^{\scriptscriptstyle\mathrm{cMF}}) = \Ham^{\scriptscriptstyle\mathrm{int}}_{\scriptscriptstyle\mathrm{cMF}}(\rhohat^{\scriptscriptstyle\mathrm{cMF}})+h \sum_i^{N_c}\PauliSigma_i^z,
\eea
where
\begin{equation}
    \Ham^{\scriptscriptstyle\mathrm{int}}_{\scriptscriptstyle\mathrm{cMF}}(\rhohat^{\scriptscriptstyle\mathrm{cMF}})=\sum_{i=1}^{N_c}\Ham^{\scriptscriptstyle\mathrm{int},i}_{\scriptscriptstyle\mathrm{cMF}}-\frac{J}{\calN_\alpha}\sum_{i,j=1}^{N_c}\frac{1}{||i-j||^\alpha}\PauliSigma_i^x\PauliSigma_j^x\,.
\end{equation}
The first term, namely $\Ham^{\scriptscriptstyle\mathrm{int},i}_{\scriptscriptstyle\mathrm{cMF}}$, is the cluster-mean-field Hamiltonian for the $i$-th site within the cluster due to the interaction with the sites of the surrounding clusters, while the second term represents the interaction between spins of the same cluster.  The former can be written as $\left(m^x_j(t) = \Tr\left\{\sigmax_j\rhohat^{\scriptscriptstyle\mathrm{cMF}}_t\right\}\right):$
\begin{equation}
\begin{split}
    \Ham^{\scriptscriptstyle\mathrm{int},i}_{\scriptscriptstyle\mathrm{cMF}}&=-2\frac{J}{\calN_\alpha}\PauliSigma_i^x\sum_{j=1}^{N_c}\sum_{\xi=1}^{\sfrac{N}{2N_c}-1}
    \bigg[\frac{1}{\underbrace{(\xi N_c+j-i)^\alpha}_{\sfrac{1}{d^\rightarrow_\xi (i,j)^\alpha}}}\\
    &+\frac{1}{\underbrace{(\xi N_c+i-j)^\alpha}_{\sfrac{1}{d^\leftarrow_\xi(i,j)^\alpha}}}\bigg]m_j^x\,.
\end{split}
\end{equation}
 Since translational invariance is not assumed within the cluster, but only among the clusters as a whole, the sum runs over $N_c$ equivalent cluster-mean-field values. Furthermore, note that
\bea\label{eq:zeta-series} \sum_{\xi=1}^{\sfrac{N}{2N_c}-1}\frac{1}{d^\rightarrow_\xi(i,j)^\alpha}&\underset{{N\rightarrow\infty}}{=}N_c^{-\alpha}\zeta(\alpha,(j-i)/N_c+1)\,,\\
\sum_{\xi=1}^{\sfrac{N}{2N_c}-1}\frac{1}{d^\leftarrow_\xi(i,j)^\alpha}&\underset{{N\rightarrow\infty}}{=}N_c^{-\alpha}\zeta(\alpha,(i-j)/N_c+1)\,,\\
\mathcal{N}_\alpha = 2\sum_{r=1}^{N/2}\dfrac{1}{r^\alpha}&\underset{{N\rightarrow\infty}}{=}2\zeta(\alpha,1)\,,
\eea
where $\zeta(\alpha,q)=\sum_{\xi=0}^{\infty}\frac{1}{(\xi+q)^\alpha}$ is the Hurwitz Zeta function. Finally, the interaction term reads:
\begin{equation}
\begin{split}
    \Ham_{\scriptscriptstyle\mathrm{int}}^{\scriptscriptstyle\mathrm{cMF},i}=-\frac{J}{\zeta(\alpha,1)}\PauliSigma_i^x\sum_{j=1}^{N_c}
    &m_j^x N_c^{-\alpha}[\zeta(\alpha,(i-j)/N_c+1)\\
    &+\zeta(\alpha,(j-i)/N_c+1)].
\end{split}
\end{equation}
The Lindblad equation in cluster mean field approximation thus becomes
\bea
    \dfrac{\rmd\rhohat^{\scriptscriptstyle\mathrm{cMF}}}{\rmd t} = -\rmi\[\Ham_{\scriptstyle\mathrm{cMF}}\(\rhohat^{\scriptscriptstyle\mathrm{cMF}}\),\rhohat^{\scriptscriptstyle\mathrm{cMF}}\]+\gamma\sum_{i=1}^{N_c}\dcal[\sigmam_i]\rhohat^{\scriptscriptstyle\mathrm{cMF}}\,,
\eea
where
\bea
    \Ham_{\scriptscriptstyle\mathrm{cMF}}\(\rhohat^{\scriptscriptstyle\mathrm{cMF}}\)=\Ham^{\scriptscriptstyle\mathrm{cMF}}_{\scriptscriptstyle\mathrm{int}}\(\rhohat^{\scriptscriptstyle\mathrm{cMF}}\)+\Ham^{ }_{\scriptscriptstyle\mathrm{ext}}\,.
\eea
Note that the series defined in Eq.~\eqref{eq:zeta-series} are convergent only for $\alpha>1$, resulting in non-trivial corrections from the intra-cluster interactions. In the contrary case of $0<\alpha<1$, one can show that the cluster interaction Hamiltonian reduces to
\bea
    \Ham^{\scriptscriptstyle\mathrm{int}}_{\scriptscriptstyle\mathrm{cMF}}\underset{\scriptscriptstyle{0<\alpha<1}}{=}-2J\bar{m}^x\sum_i\sigmahat^x_i\,,\quad~\bar{m}^x \equiv \frac{1}{N_c} \sum_{j=1}^{N_c} m_j^x\,,
\eea
which completely decouples, as expected from the exactness of the single-site mean field theory in the strong-long-range regime~\cite{mattesLongRangeInteractingSystems2025}.

This approach allows to study the regime of shorter-range interactions by checking convergence with increasing $N_c$. As shown in Fig.~\ref{fig:cluster-mf-compare}, for sufficiently long-range interaction $\alpha\lesssim 2$, the converging results imply the survival of the ferromagnetic phase, as opposed to the case with $\alpha>2$, where the region with nonzero magnetization $|\langle\sigmahat^x\rangle|$ shrinks with $N_c$, suggesting that the ferromagnetic phase might vanish in exact solution.

\section{Derivation of the cumulant equations}\label{app:cum}

In Eq.~\eqref{eq:cum-tilt}, the second cumulants $v_{kl}^{\alpha\beta}$ are expectations of time-dependent operators by definition, of the form $\delhat_A\delhat_B=\Ahat\Bhat-\langle\Ahat\rangle\langle\Bhat\rangle$ for some time-independent operators $\Ahat$ and $\Bhat$, and their dynamics can be derived with a convenient identity that we prove in this section. First, it is easy to show that the normalized adjoint tilted Lindbladian is linear,
\bea
    \tilde\Lcal^\ddagger_\chivec ( \hat A + \lambda \hat B ) = \tilde\Lcal^\ddagger_\chivec \hat A + \lambda \tilde\Lcal^\ddagger_\chivec\hat B\,,
\eea
for any operators $\hat A,\hat B$ and $\lambda\in \mathbb{C}$, and that
$\langle \tilde\Lcal^\ddagger_\chivec \idhat \rangle_\chivec = 0$. Recalling the definition of the fluctuation operator $\delhat_A = \Ahat - \langle\Ahat\rangle_\chivec$ (and idem for $\Bhat$), we have,
\bea\label{eq:dv-cum}
    \dfrac{\rmd}{\rmd t}\langle\delhat_A\delhat_B\rangle_\chivec &= \dfrac{\rmd}{\rmd t} ( \langle\Ahat\Bhat\rangle_\chivec - \langle\Ahat\rangle_\chivec\langle\Bhat\rangle_\chivec )\\
    &= \langle  \tilde\Lcal_\chivec^\ddagger(\Ahat\Bhat) \rangle_\chivec - \langle  \tilde\Lcal_\chivec^\ddagger\Ahat \rangle_\chivec\langle\Bhat\rangle_\chivec - \langle\Ahat\rangle_\chivec\langle\tilde\Lcal_\chivec^\ddagger\Bhat\rangle_\chivec\\
    &= \langle  \tilde\Lcal_\chivec^\ddagger( \Ahat\Bhat - \Ahat\langle\Bhat\rangle_\chivec - \langle\Ahat\rangle_\chivec\Bhat  ) \rangle_\chivec \\
    &= \langle \tilde\Lcal_\chivec^\ddagger( \Ahat\Bhat - \Ahat\langle\Bhat\rangle_\chivec - \langle\Ahat\rangle_\chivec\Bhat + \langle\Ahat\rangle_\chivec\langle\Bhat\rangle_\chivec )\rangle_\chivec
    \\
    &= \langle\tilde\Lcal_\chivec^\ddagger(\delhat_A\delhat_B)\rangle_\chivec\,.
\eea
The same identity applies to the standard Liouvillian, as a special case of $\chivec=\Vec{0}$.

For the dissipative spin model discussed in the main text, these equations are explicitly written as follows,
\bea
    \dfrac{\rmd}{\rmd t}\cscr =& 2\gamma \cscr \sum_j( \rme^{\rmi\chi_j}-1 )(c^z_j + 1)\,,
\eea
\bea
    \dfrac{\rmd}{\rmd t} c^\alpha_k =& -2h \epsilon_{z\alpha\gamma}c^\gamma_k + 4\epsilon_{x\alpha\gamma}\sum_{n\neq k}J_{nk}\left( v^{x\gamma}_{nk} + c^x_n c^\gamma_k \right)\\
    & -2\gamma\left[ \rme^{\rmi\chi_k} \delta^{\alpha z}(c^z_k+1) + \delta^{\alpha z} + c^\alpha_k \right]\\
    &+ 2\gamma\sum_{j\neq k}(\rme^{\rmi\chi_j}-1)\left( v^{\alpha z}_{kj} + c^\alpha_k c^z_j  + c^\alpha_k\right)\\
    &- c^\alpha_k  2\gamma \sum_j( \rme^{\rmi\chi_j}-1 )(c^z_j + 1) \,,
\eea
\bea
    \dfrac{\rmd}{\rmd t}v^{\alpha\beta}_{kl} =& -2h( \epsilon_{z\alpha\gamma}v^{\gamma\beta}_{kl} + \epsilon_{z\beta\gamma}v^{\alpha\gamma}_{kl} )\\
    &+ 4\epsilon_{x\alpha\gamma}\sum_{n\neq k,l} J_{nk}( c^\gamma_k v^{x\beta}_{nl} + c^x_n v^{\gamma\beta}_{kl} )\\
    &+4\epsilon_{x\beta\gamma}\sum_{n\neq k,l}J_{nl}( c^\gamma_l v^{x\alpha}_{nk} + c^x_n v^{\gamma\alpha}_{lk} )\\
    &+ 4J_{kl}\epsilon_{x\alpha\gamma}\left[ \delta^{\beta x}c^\gamma_k - c^\beta_l\left(   c^\gamma_k c^x_l + v^{\gamma x}_{kl} \right) \right]\\
    &+ 4J_{kl}\epsilon_{x\beta\gamma}\left[ \delta^{\alpha x}c^\gamma_l - c^\alpha_k\left(   c^\gamma_l c^x_k + v^{\gamma x}_{lk} \right) \right]\\
    &+2\gamma\sum_{n\neq k,l}(\rme^{\rmi\chi_n}-1)\left( c^z_n+1  \right) v^{\alpha\beta}_{kl}\\
    &- 2\gamma\left[ (\delta^{\alpha z}\rme^{\rmi\chi_k}+1)v^{\alpha\beta}_{kl} + c^\alpha_k(\rme^{\rmi\chi_k}-1)v^{z\beta}_{kl} \right]\\
    &- 2\gamma\left[ (\delta^{\beta z}\rme^{\rmi\chi_l}+1)v^{\alpha\beta}_{kl} + c^\beta_l(\rme^{\rmi\chi_l}-1)v^{\alpha z}_{kl} \right]\\
    &- v^{\alpha\beta}_{kl}  2\gamma \sum_j( \rme^{\rmi\chi_j}-1 )(c^z_j + 1) \,,
\eea
where $J_{ij} = J/( r^\alpha_{ij} \mathcal{N}_\alpha )$ denotes the coupling between two sites $i$ and $j$ in the generic case.

\bibliography{apssamp}

\end{document}